\documentstyle[prl,aps,preprint]{revtex}
\begin{document}

\title{Stability of optically-active charged excitons 
in quasi-two dimensional systems}
\author{James R. Chapman, Neil F. Johnson, 
V. Nikos Nicopoulos }
\address{ Department of Physics, Clarendon Laboratory, 
Oxford University,
Oxford OX1 3PU, England}
\maketitle

\begin{abstract}

A negatively charged quasi-two dimensional exciton ($X^-$) is 
solved exactly numerically
 in the presence of a uniform perpendicular B-field. 
Various quasi-two 
dimensional geometries are studied. The charge 
distribution of the $X^-$ parallel
 to the B-field is found to be crucial in determining
 the stability of the optically-active $X^-$ and 
hence its photoluminescence (PL) signature. The
 theory provides a quantitative
 explanation of recent experimental results obtained 
for a GaAs quantum well. 
Effects are found which cannot be described within 
a lowest Landau level 
approximation. \\
PACS: 78.20.Ls 78.66.-w 73.20.Dx

\end{abstract}

\newpage

Excitonic effects are used to explain the majority 
of discrete lines appearing
 in the interband optical spectra obtained from
 semiconductors. However new
 lines have been seen recently which cannot be
 explained by excitons, either
 intrinsic or impurity-bound (see eg Ref.
 ~\cite{Finkelstein2} ~\cite{Shields2}). 
It has been proposed that these discrete 
lines are associated with the formation of a 
charged electron-hole complex.
 The stability of charged excitons was first 
predicted for three-dimensional 
(3D) systems by Lampert ~\cite{Lampert}.  
This was subsequently studied
 theoretically at zero B-field in both 3D~\cite{Munschy} 
and two-dimensions
 (2D)~\cite{Stebe}. In both cases the charged exciton was
 found to be stable.
The 2D charged exciton binding energies are about ten 
times larger than than those in 3D
 due to the extra confinement~\cite{Stebe} thereby
 facilitating their experimental
 observation. Negatively charged excitons, $X^-=2e+1h$,
 were first observed in
 CdTe/CdZnTe quantum wells ~\cite{Kheng1} 
~\cite{Kheng2} and were later
 seen in GaAs/GaAlAs quantum wells~\cite{Finkelstein1}
 ~\cite{Finkelstein2}
 ~\cite{Shields1} ~\cite{Shields2} ~\cite{Shields3}.  Positively 
charged excitons, $X^+=1e+2h$,
 have also been seen~\cite{Finkelstein2}. The charged excitons
 have been observed
 experimentally using polarized spectroscopy, in
 particular photoluminescence
 (PL) with a uniform magnetic field B applied
 perpendicular to the quantum 
well plane. $X^-$ stability  requires a two-dimensional
 electron gas (2DEG) 
of intermediate electron density which is typically
 near the metal-insulator
 transition, the reduced screening effect of the other
 electrons helping to
 stabilise the $X^-$~\cite{Finkelstein1}. In the low
 density limit the exciton
 ($X$) dominates the PL spectrum as expected.
 At high densities and low 
B-fields, a Fermi-edge-singularity (FES) is seen. 
The FES moves smoothly
 into the $X^-$ line as the electron density is
 decreased~\cite{Finkelstein2}
 ~\cite{Shields1} and develops into sharp $X$ 
and $X^-$ peaks with increasing
 B-field ~\cite{Kheng2}. These results suggest 
that a study of the $X^-$ 
is crucial for an understanding of the optical 
properties of 2DEGs at finite
 B-fields, as was claimed recently in Ref.~\cite{Finkelstein2}.

In this paper an accurate numerical solution is
 presented for a quasi-2D
 $X^-$ consisting of two electrons and a 
hole situated in a weak 
 in-plane confinement potential and a 
perpendicular uniform B-field. 
The $X^-$ properties are studied over a wide
 range of experimentally 
applicable B-fields. Various relevant quasi-2D
 geometries are considered,
 each having a different charge distribution
 parallel to the B-field, ie. 
perpendicular to  the 2DEG. These geometries
 are as follows: \\ (i) electrons 
and hole are strictly 2D but move on separate
 parallel planes (biplanar
 geometry), \\ (ii) electrons and hole occupy the 
same plane but  have 
their charge smeared in a  rod-like distribution
 along the perpendicular
 direction (rod geometry),\\ (iii) electrons and
 hole are strictly 2D and 
occupy the same plane (coplanar geometry). \\  
The stability of the optically-active $X^-$
 and hence its PL spectrum are found to vary
 drastically depending on
 the geometry used. It is shown that recent
 experimental PL spectra obtained
 from quantum wells can only be explained 
using the rod geometry {\em and} 
higher Landau level mixing. A previous theoretical 
study only considered a coplanar
geometry at a fixed B-field ~\cite{Hawrylak}.

 The Hamiltonian for the $X^-$ problem is:

\begin{center}
\begin{eqnarray}

 {\mathcal{H}} &=& {1 \over 2m_{h}}\left[ 
{\bf p_{h}} - {e \over c}{\bf A}
({\bf r_{h}})\right]^{2} + {1 \over 2}m_{h} 
{\omega}_{h}^{2}{\bf r_{h}^{2}} 
+g_h \mu S^{z}_{h} \nonumber \\ &+& 
\sum_{i=1}^2 \left\{ {1 \over 2m_e}
 \left[ {\bf p_{e,i}} + {e \over c} {\bf A} 
({\bf r_{e,i}}) \right]^2 + {1 \over 2} 
m_e {\omega}^2_e {\bf r_{e,i}^2} +
 g_e \mu S_{e,i}^z - V_{eh}(|{\bf r_{e,i}}
 - {\bf r_h}|) \right\} \nonumber \\ 
&+& V_{ee}(|{\bf r_{e,1}} - {\bf r_{e,2}}|)
\end{eqnarray}
\end{center}
where $m_e = 0.067 m_0$ and $m_h = 
0.475 m_0$ are the masses of the
  conduction electrons and heavy holes respectively
 for GaAs. In order to
 facilitate comparison with a quasi-2D electron 
gas the parabolic confinement
 potentials, $\hbar \omega_e = 1.75$meV and
 $\hbar \omega_h = 0.25$meV, 
are chosen to be weak compared with the
 cyclotron energies ($\hbar
 (\omega_c)_e = 17.2$meV and $\hbar (\omega_c)_h
 = 2.4$meV at B=10T).
 Also ${\omega_e / \omega_h} = {m_h / m_e}$
 ensuring that the electron
 and hole wavefunctions have equal characteristic
 lengths, $l_{e,h}^2 = 
{\hbar / (m_{e,h} \tilde{\omega}_{e,h})} = l^2$
 where $\tilde{\omega}_{e,h}^2
 = ((\omega_c)_{e,h}^2 + 4 {\omega}_{e,h}^2)$, as
 is the case in the
 limit of zero confinement.

The interaction potentials are Coulombic. 
In order to obtain a
 tractable model it is assumed that the single
 particle wavefunctions
 separate: ${\Psi}_e = {\phi}_e(z_e){\psi}_e({\bf r_e})$
 and ${\Psi}_h = 
{\phi}_h(z_h){\psi}_h({\bf r_h})$.  A specific form
 for ${\phi}_{e,h}$ is chosen,
 thereby freezing out the z-motion and yielding
 a quasi-two dimensional
 model. For the biplanar geometry:
\begin{center}
\begin{equation}
|{\phi}_e(z_e)|^2 = \delta (z_e - 0) \ \mbox{ ; }\ 
|{\phi}_h(z_h)|^2 = \delta (z_h - d)
\end{equation}
\end{center}
while for the rod geometry:
\begin{center}
\begin{equation}
|{\phi}_{e,h}(z_{e,h})|^2 = \left\{ \begin{array}{ll}
                          {1 \over L}  & \mbox{if ${-L \over 2} 
< z_{e,h} < {L \over 2}$}  \\
                          0  &  \mbox{otherwise}
                        \end{array}
                       \right.
\end{equation}
\end{center}
The coplanar geometry is the limit of the 
previous two where $d$ 
and $L$ tend to zero.\\
The coplanar and rod models retain a 
singularity in $V_{e,h}$ 
at ${\bf r_e=r_h}$ which the biplanar 
model loses. The differences in 
the forms of $|\phi (z)|^2$ lead to significantly
 different PL spectra.

 The effects of the B-field and the Coulomb
 interaction compete: the 
expansion parameters are given by ${a_{e,h} 
/ l}$ where $a_{e,h} = \epsilon
 \hbar^2/m_{e,h}e^2$ is the Bohr radius for 
the electron and hole respectively.
 Taking $B \sim 8$T and $\epsilon_r = 12.53$
 yields (GaAs) ${a_h / l} =0.2$ 
and ${a_e / l} = 1.1$, thereby rendering any 
perturbation approach unreliable. 
For this reason the $X^-$ Hamiltonian is diagonalized 
exactly numerically.
Rotational symmetry about the B-field leads 
to conservation of total angular
 momentum, M. The symmetric gauge is hence chosen, 
${\bf A}({\bf r})=({\bf B \times r})/2$,
 giving  single particle states characterised by 
the Landau level ($n$) and 
azimuthal angular momentum ($m$). Angular 
momentum 
conservation makes the Hamiltonian block 
diagonal in this basis, each block 
being characterised by M.
The spin and spatial components of the 
wavefunction factorise, the spin
 states of the electrons being either singlet 
or triplet. In order to preserve
 the correct electron antisymmetry the triplet
 (singlet) states have a spatial 
component which contains antisymmetric
 (symmetric) combinations of the
 single particle electron wavefunctions.

The Hamiltonian is diagonalized in the 
basis of definite M, S and 
$\mbox{S}_z$. Since ${(a_{e,h} / l)}  > 1$ 
 higher Landau levels must be
 included in the basis. The actual basis 
used has 6 hole Landau levels, 2 electron
 Landau levels and 20 angular momentum 
states per Landau level 
(cf.~\cite{Hawrylak}). A larger number of 
hole Landau levels are used because 
of the small heavy-hole cyclotron energy. 
A finite number of angular momentum
 states are used due to the weak confinement 
which removes the degeneracy
 of the Landau level. From comparison 
with experiment (discussed later in this 
paper) the authors believe that this basis
 captures the essential physics of the
 $X^-$. The matrix elements are evaluated 
following a method outlined in 
~\cite{Sham}. In the coplanar limit ($L,d 
\rightarrow 0$) analytic results are
 obtained for the matrix elements ~\cite{Halonen}
 ~\cite{Hawrylak}. Away from
 this limit, numerical integration is employed.
 Two completely different methods
 for evaluating the matrix elements have
 been used to check for numerical error
 in the integration routines.
For comparison purposes the exciton lines
 are also calculated in a manner
 identical to that used for the $X^-$.

Both excitons ($X$) and $X^-$ undergo recombination
 giving rise to a PL spectrum.
 It is assumed here, as in ~\cite{Shields2},
 that recombination is with the 
$\pm 3/2 $ heavy holes. The oscillator 
strengths
 for the $X$ and $X^-$ PL lines are 
 $\langle 0| \hat{L} |X \rangle$ and  
$\langle e| \hat{L} |X^-\rangle$ respectively,
  where  $\hat{L} = \int {\hat{\psi}_e}
({\bf r}) {\hat{\psi}_h}({\bf r}) d^2{\bf r} $.
 This means that only excitons
 with orbital angular momentum M=0 will 
recombine. For both $X$ and 
$X^-$ the necessary change in the system's
 angular momentum is taken up
 by the atomic part of the wavefunction
 giving rise to two lines, $\sigma^+$ 
and $\sigma^-$, which are Zeeman split. 
In the limit that the basis is restricted to 
the lowest Landau level and setting 
$V_{eh} = V_{ee}$, there is a well-documented
 hidden symmetry~\cite{MacDonald}
 ~\cite{Apalkov}. This forces the triplet 
$X^-$ $(X^-_t)$ PL energy to be identical
 with that of the lowest exciton $(X)$ line. 
In this paper the symmetry is broken
 by the weak confinement and more 
importantly by the inclusion of higher
 Landau levels. For the case of a biplanar
 z-geometry the interactions $V_{ee}$
 and $V_{eh}$ are no longer equal, thus 
further breaking the hidden symmetry. 
The reliability of the calculation has been
 verified  by reproducing known results
 as follows: 
(a) in the limit of zero confinement and
 using the rod geometry ($V_{ee}=V_{eh}$)
 the hidden symmetry result is recovered, 
(b) the B=0 confined 2-electron results of 
~\cite{Merkt} are reproduced, 
(c) results found in a previous $X^-$ study 
~\cite{Hawrylak} are obtained.

The theoretical PL spectra resulting from the 
coplanar, biplanar and rod geometries
 are now discussed.  
When a strictly coplanar geometry is used
 both the optically-active singlet ($X^-_s$) 
and triplet ($X^-_t$) states
 are stable with respect to the decompositon 
$X^- \rightarrow X + e$. This 
stability persists over a calculated range in
 B-field of at least $6 \rightarrow 14$T. 
However, the agreement with recent experimental
 quantum well data (see Fig 1a) 
is not good since the calculated  $X - X_s^-$
 splitting is too large (3.8meV at B=8T)
 and there is no strong increase of the PL 
energies with B.
Using the biplanar geometry the PL spectrum
 is found to be very sensitive to
 changes in the plane separation, $d$. For 
an $X^-$ with zero total angular 
momentum, M=0, the optically-active
 singlet and triplet species
 are stable when $d=0$. Increasing $d$ from 
zero causes the singlet $(X^-_s)$ to unbind ($d
 \sim 0.5l$) with the triplet $(X^-_t)$ unbinding
 soon after ($d \sim l$). Similar unbinding
 occurs for the optically-active M$\neq$0 $X^-$
 studied. Thus the biplanar geometry PL spectra
 differ strongly from the experimental
 quantum well PL spectra  (Fig 1a). These results 
demonstrate that quasi-2D systems 
which approximate to a biplanar system (eg 
heterojunctions) with $d > l$ are
 most unlikely to exhibit PL effects due to charged excitons.

The experimental PL spectrum of Shields
 et al. for a GaAs quantum 
well~\cite{Shields2}  shown in Fig 1a  is 
typical of those found in the 
literature. The exciton ($X$), singlet $X^-$
($X^-_s$), and triplet $X^-$ ($X^-_t$)
 are marked in accordance with the 
experimental assignments of Ref. ~\cite{Shields2}
 (for clarity, only the $\sigma^-$ lines are
 shown in all figures in this paper).
The theoretical lines shown in Fig 1b are 
calculated using the rod geometry
 with a rod length $L=$225\AA \  and total 
angular momentum M=0. For the 
range of B-fields shown in Fig 1, $L$ is 
less than the width of the $X^-$ in the 
plane thus validating the assumption 
that the $X^-$ is quasi-2D. The $L$ value
 is also consistent with the width of the 
envelope function in the experimental 
quantum well used by 
Shields et al ~\cite{Shields2}. Figure 1
 shows that the theoretical lines exhibit
 many of the quantitative features of the
 experimental spectrum. The energy
 of the calculated PL lines increases with
 B-field in a manner closely matching
 that of the experimental lines. This is a 
non-trivial agreement since the slope
 is {\em not} the result of simply adding
 the single-particle cyclotron energies.
 The $X-X_s^-$ splitting ($\sim 1.4$meV
 at 8T) is close to the experimental
 value ($\sim 1.7$meV at 8T). With
 increasing B-field the triplet line ($X_t^-$) 
becomes more separated from the exciton
 line ($X$) hence $X^-_t$ becomes 
increasingly stable. However, $X^-_t$
 only becomes stable at a finite B-field 
(5T theoretically compared with 2.5T
 experimentally). The ordering of the PL 
lines in energy is very stable to changes in $L$. 
 The quantitative theory discussed 
here thus allows a definite assignment of 
the PL lines which is in agreement 
with that proposed for the original
 experimental spectra ~\cite{Shields2}. The 
slight differences between the theory and 
experiment in Fig 1 are to be expected
 since the model is necessarily a simple,
 quasi-2D representation 
of a complex 3D system. 

Recombination from charged excitons in the rod geometry
 with total angular momentum 
M$<$0 should also be considered as they 
give PL lines of similar energies to 
M=0 $X^-$. The PL spectra for a M=$-$2 
$X^-$ at B=10T are calculated both with 
and without an in-plane confinement 
and are shown in Figs 2a and 2b respectively.
 As the in-plane confinement is removed 
the lowest triplet state has its oscillator 
strength drastically lowered to the benefit 
of the first excited triplet state (cf Figs
 2a and 2b). The PL spectra in the zero confinement 
limit agree well with 
the experimental PL spectra in Ref.~\cite{Shields2}. 
The samples used in Ref.~\cite{Shields2}  are reported to have 
a thick undoped spacer layer between the donor atoms
 and the well. This minimizes the spatial confinement 
in the plane of the well, giving mobilities which demonstrate
 a high degree of homogeneity and allowing comparison
 with calculations in the zero confinement limit.  
 Fig 2c shows the PL spectrum of the M=0
 $X^-$ for B=10T with the weak in-plane 
confinement. As the in-plane confinement
 is removed (Fig 2d) the PL spectrum is almost 
unaffected. If the basis is restricted
 to the lowest Landau level, the hidden 
symmetry result (triplet and exciton PL lines
 coincident) is obtained when the in-plane
 confinement is removed (see inset in Fig 2).
 The fact that the splitting is {\em retained} 
in Figs 2b and d implies that higher
 Landau levels are crucially important in
 allowing the optically-active $X^-_t$ wavefunctions to 
form configurations stable against 
the decomposition $X^-_t \rightarrow
 X+e^-$: this breaks the hidden symmetry 
and causes the $X^-_t$ PL line to
 be seen. In fact the energy levels of the $X^-$
 and exciton {\em all} change
significantly when the basis is restricted to the 
lowest Landau level.

To summarize, a quasi-two dimensional $X^-$
 has been solved exactly
 numerically in the presence of a uniform 
perpendicular magnetic field and 
a  weak in-plane parabolic confinement. The
 diagonalization basis contained
 higher Landau levels giving rise to results 
which differ significantly from those
 obtained within a lowest Landau level approximation. 
The quasi-2D coplanar, 
biplanar and rod geometries studied  were 
found to yield significantly different
 PL spectra. Of all the geometries considered,
 the rod model approximates 
closest to a quantum well and was able to produce
 good quantitative agreement
 with recent experimental quantum well spectra.
 The biplanar model predicts
 an instability of optically-active $X^-$ with increasing $d$: 
this is consistent with the apparent
 failure to observe charged excitons in experimental 
quasi-2D systems other than quantum wells.

The authors would like to thank Darren 
Leonard for illuminating discussions and J. J. Palacios
for a helpful comment.
 Funding was provided by EPSRC through
 a studentship (J.R.C.) and a Research
 Grant No. GR/K 15619 (N.F.J. and V.N.N.).

\newpage \centerline{\bf Figure Captions}

\bigskip

\noindent Figure 1: Variation of $X^-$
 and exciton $X$ PL spectra ($\sigma^-$)
with B-field. (a) Experimental; adapted from
 ~\cite{Shields2} for a 300 \AA \ GaAs 
quantum well. (b) Theory; zero of energy 
shifted to  include the band gap. Ranges
 are chosen to highlight the agreement of 
trends in the experimental and calculated
 spectra. Parameters given in text. Lines 
are guides to the eye.
\bigskip

\noindent Figure 2:  PL spectrum ($\sigma^-$)
 at B=10T,
 $L=$225\AA \ for (a) M=$-$2, weak 
in-plane confinement (b) M=$-$2, zero in-plane
 confinement limit (c) M=0, weak 
in-plane confinement (d) M=0, zero in-plane 
confinement limit. Parameters given 
in text.
 Inset shows $X^-_t$ and $X$ lines calculated
 in the lowest Landau level
 (LLL) with zero in-plane confinement. A
 gaussian line-broadening has been 
introduced throughout. 
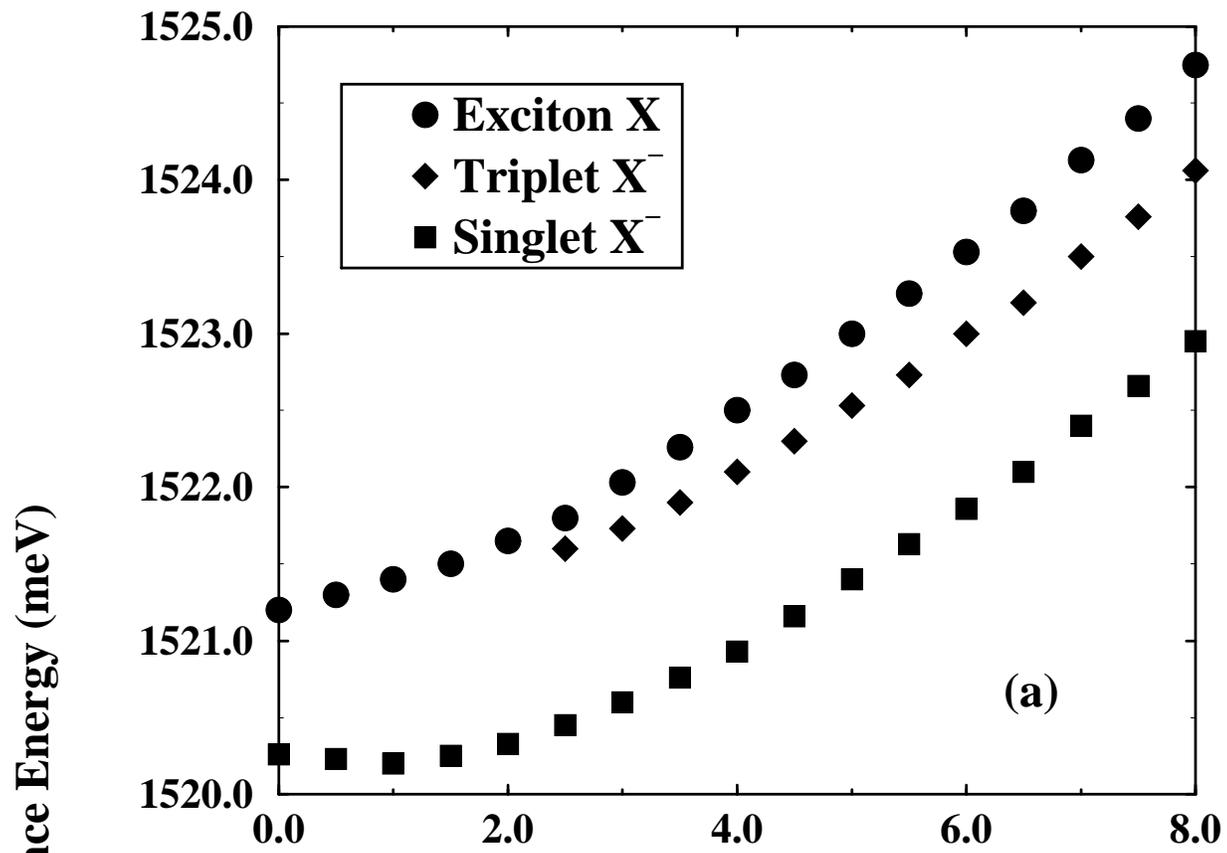
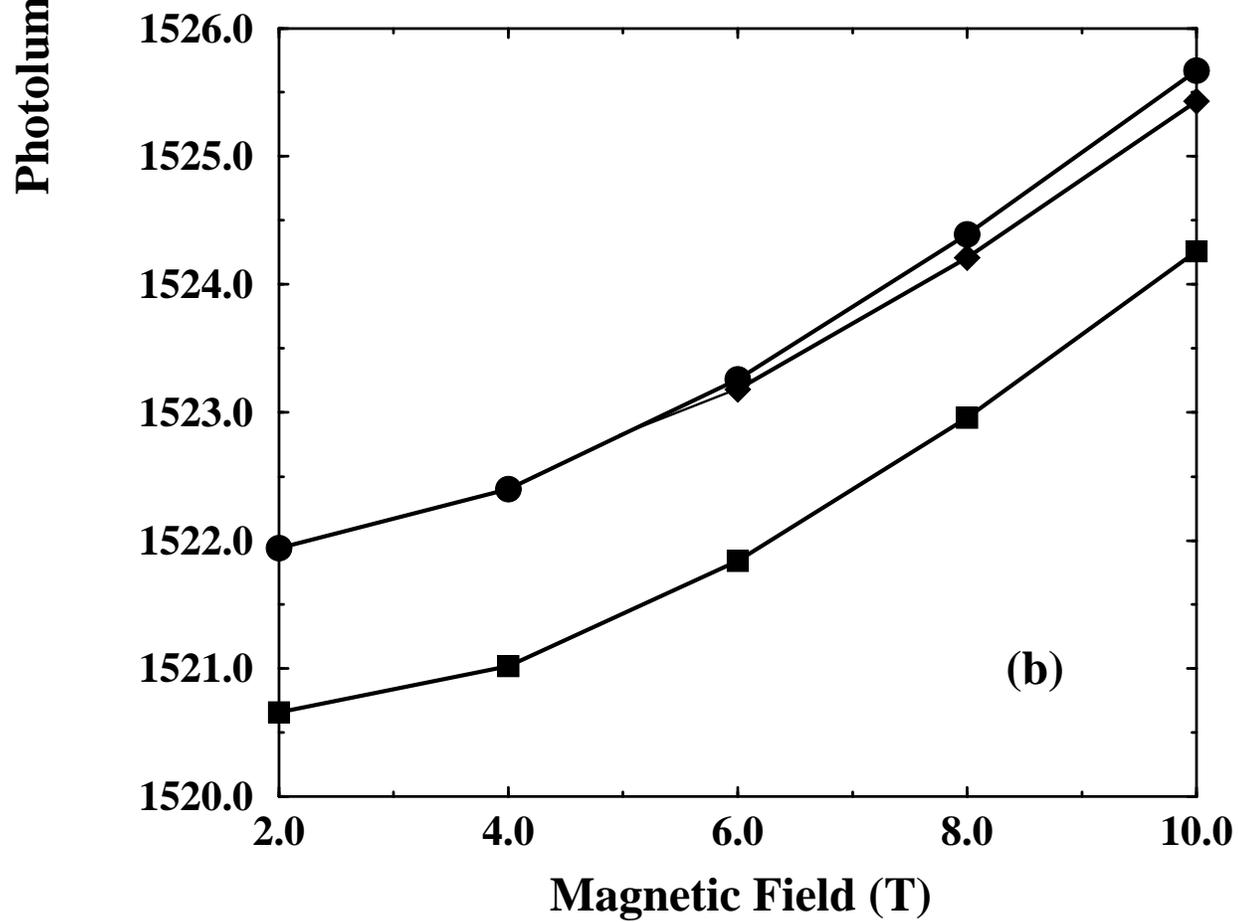
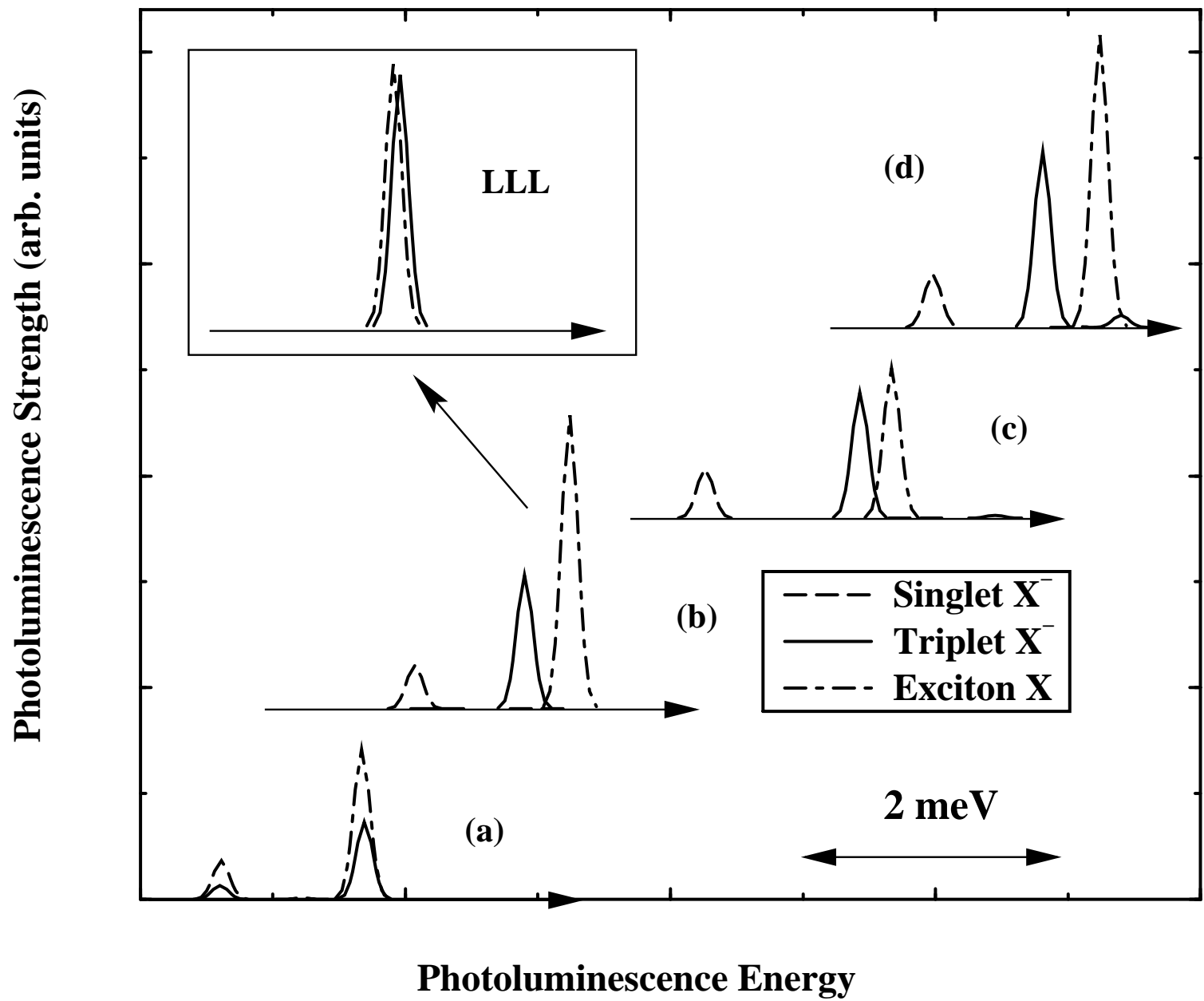
\end{document}